%% file: mainpaper.tex
\documentclass[runningheads]{llncs}
\usepackage{graphicx}
\usepackage{multirow}
\usepackage{hyperref}
\usepackage{multicol}
\usepackage{marvosym}
\usepackage{cite}
\hypersetup{
    colorlinks=true,
    linkcolor=blue,
    citecolor=blue,
    urlcolor=blue,
    }
\begin{document}
\title{Simplified TinyBERT: Knowledge Distillation for Document Retrieval}
%
\author{Xuanang Chen\inst{1,2}\textsuperscript{(\Letter)} \and
Ben He\inst{1,2}\textsuperscript{(\Letter)} \and
Kai Hui\inst{3}\thanks{This work has been done before joining Amazon.} \and
Le Sun\inst{2} \and
Yingfei Sun\inst{1}\textsuperscript{(\Letter)}}

\authorrunning{X. Chen et al.}

\institute{University of Chinese Academy of Sciences, Beijing, China 
\and Institute of Software, Chinese Academy of Sciences, Beijing, China 
\email{chenxuanang19@mails.ucas.ac.cn
\\ \{benhe, yfsun\}@ucas.ac.cn, sunle@iscas.ac.cn}
\and Amazon Alexa, Berlin, Germany \\ \email{kaihuibj@amazon.com}
}

\maketitle              
\begin{abstract}
Despite the effectiveness of utilizing the BERT model for document ranking, the high computational cost of such approaches limits their uses. 
To this end, this paper first empirically investigates the effectiveness of two knowledge  distillation models on the document ranking task. 
In addition, on top of the recently proposed TinyBERT model, two simplifications are proposed.
Evaluations on two different and widely-used benchmarks demonstrate that Simplified TinyBERT with the proposed simplifications not only boosts TinyBERT, but also significantly outperforms BERT-Base when providing 15$\times$ speedup.


\end{abstract}

\input{intro}

\input{background}

\input{method}

\input{expResults}

\input{conclusions}

%
%
%
\bibliographystyle{splncs04}
\bibliography{mainpaper}

\end{document}

%% file: intro.tex
\section{Introduction}\label{sec:intro}
Contextual pre-trained model, like BERT~\cite{DBLP:bert18}, demonstrates its effectiveness in ranking tasks~\cite{DBLP:Dai19,DBLP:nyu19,DBLP:Yang19}. However, the vast number of parameters in BERT make it expensive or even infeasible for serving~\cite{DBLP:conf/sigir/HofstatterH19,DBLP:CEDR}, which is especially important when the model is used to re-rank thousands of search results. In the meantime, studies~\cite{DBLP:tinybert,DBLP:DistilBERT,DBLP:SunCGL19,DBLP:MobileBERT,DBLP:lin,DBLP:wellread,DBLP:minilm} have demonstrated that knowledge distillation (KD) can be used to learn smaller BERT models without compromising effectiveness too much, wherein a full-sized BERT model, like BERT-Base, is used as the teacher model and a small student model is trained to imitate it. More specifically, TinyBERT~\cite{DBLP:tinybert} is proposed to distill on both prediction layer and intermediate layers in a two-stage distillation method, and has achieved effectiveness that is close to the teacher model on multiple NLP tasks. However, it is unclear whether such distillation models are still effective on the document ranking task.

To bridge this gap, in this work, we first investigate the uses of the standard knowledge distillation model~\cite{DBLP:HintonVD15} and the more recent TinyBERT~\cite{DBLP:tinybert} on the document ranking task. In addition, we propose two simplifications for TinyBERT, hoping to further improve the effectiveness of the distilled ranking models. To this end, on the document ranking task in MS MARCO~\cite{DBLP:msmarco16} and TREC 2019 DL Track \cite{DBLP:TREC2019DL},
we demonstrate the potentials in employing knowledge distillation for document retrieval, and also confirm the superior effectiveness of the proposed Simplified TinyBERT which will be described in Section~\ref{sec:modified_tinybert}. 

The contributions of this work are twofold. (1) To the best of our knowledge, this is the first effort to employ knowledge distillation for the document ranking task, by empirically investigating the effectiveness of standard knowledge distillation model~\cite{DBLP:HintonVD15} and TinyBERT~\cite{DBLP:tinybert} on two document ranking benchmarks; and, (2) Two simple but effective modifications have been proposed on top of TinyBERT. The student model distilled with the proposed Simplified TinyBERT not only can boost TinyBERT, but also significantly outperform BERT-Base when providing 15$\times$ speedup. The source code is available at \href{https://github.com/cxa-unique/Simplified-TinyBERT}{https://github.com/cxa-unique/Simplified-TinyBERT}.

%% file: background.tex
\section{Background}\label{sec:background}
\textbf{Passage-level BERT-based Document Re-ranking.} Given a query and a document, the document is first split into overlapping passages, before a BERT model consumes the concatenation of query and passage through multiple transformer layers, and ultimately generates a score to indicate the relevance of the passage relative to the query.
After that, the score of a document can be produced by its best passage (BERT-MaxP~\cite{DBLP:Dai19}), which is used to re-rank the documents.

\textbf{Knowledge Distillation (KD).}
Due to the expensive computation cost of BERT during inference, some KD methods on BERT have been proposed, such as DistilBERT~\cite{DBLP:DistilBERT}, BERT-PKD~\cite{DBLP:SunCGL19}, TinyBERT~\cite{DBLP:tinybert}, and MiniLM~\cite{DBLP:minilm}. Early KD method~\cite{DBLP:HintonVD15} relies on the soft label from the teacher model, wherein a loss function is designed to make the student model directly simulate the output of the teacher model.
In the meantime, the actual annotations are also considered in the loss function as in~\cite{DBLP:HintonVD15,DBLP:SunCGL19,DBLP:lin}.
These two kinds of cross-entropy losses are coined as the soft loss, denoted as $\mathcal{L}_{soft}$, and the hard loss, denoted as $\mathcal{L}_{hard}$, respectively.

In \textbf{TinyBERT}~\cite{DBLP:tinybert}, the pre-training and fine-tuning knowledge is distilled from a pre-trained BERT and the fine-tuned BERT on target tasks in the general stage and the task-specific stage, respectively. It employs three MSE losses to make the student model learn from three kinds of internal weights of the teacher model, namely, the attention weights, the hidden state weights, and the embedding weights in different layers, which are correspondingly denoted as $\mathcal{L}_{attn}$, $\mathcal{L}_{hidn}$ and $\mathcal{L}_{emb}$, in addition to $\mathcal{L}_{soft}$.
The intermediate layers are distilled with $\mathcal{L}_{attn}$, $\mathcal{L}_{hidn}$ and $\mathcal{L}_{emb}$ in both stages, and the prediction layer is distilled with $\mathcal{L}_{soft}$ only in the task-specific stage.

%% file: method.tex
\section{Simplified TinyBERT for Ranking}\label{sec:modified_tinybert}
In this Section, we propose two simplifications for the TinyBERT model, hoping to achieve better performance on the document ranking task.
\subsection{Method}

\textbf{Merge two steps in the task-specific stage into one step.}
As described in Section~\ref{sec:background}, TinyBERT involves two stages, and there are two steps in the second stage, wherein the training process is time-consuming. Through our empirical investigations, however, we find that the two steps could be merged into one step by simply optimizing all losses at once as described in Equation~(\ref{equ:one_soft}).
This simplification not only brings down the training time, but also boosts the ranking performance as can be seen in Table~\ref{tab:results}.
This implies that the student model could learn the prediction layer together with the intermediate layers more effectively.
Actually, we also find that one could further simplify TinyBERT distillation process by merging two stages into one, namely, employing a pre-trained BERT model, if available, and using its first $k$ layers to initialize the student model in place of the general distillation stage. For example, the student model coined as $\rm L6\_H768$ in Table~\ref{tab:results} could also be distilled with only one stage by initializing the student model with the first six layers from BERT-Base, without compromising performance. 
We will leave further investigations on this part in future work. 
\begin{equation}\label{equ:one_soft}
\mathcal{L}=\mathcal{L}_{attn}+\mathcal{L}_{hidn}+\mathcal{L}_{emb}+\mathcal{L}_{soft}
\end{equation}

\textbf{Include hard label in the loss function.} Inspired by existing models from~\cite{DBLP:HintonVD15,DBLP:lin,DBLP:SunCGL19}, 
we conjecture that the hard labels could help to distinguish the relevant and non-relevant documents better.
Therefore, we include the hard loss during distillation by adding it into Equation~(\ref{equ:one_soft}), ending up with Equation~(\ref{equ:soft_hard}). 
\begin{equation}\label{equ:soft_hard}
\mathcal{L}_{h}=\mathcal{L}_{attn}+\mathcal{L}_{hidn}+\mathcal{L}_{emb}+\mathcal{L}_{soft} +
\mathcal{L}_{hard}
\end{equation}

\subsection{Implementation Details}\label{sec:implementation}

\textbf{Use BERT-Base as the teacher model.}
In BERT-PKD~\cite{DBLP:SunCGL19},
it has been demonstrated that the uses of BERT-Base model are as effective as
when using the three-times larger BERT-Large model.
Thereby, we employ BERT-Base as the teacher model in this work,
wherein the checkpoint which is trained on MS MARCO passage dataset from~\cite{DBLP:nyu19} is used to initialize the model as in~\cite{DBLP:Birch}.
The teacher model can be further fine-tuned on MS MARCO document dataset, but is omitted in our experiments, as both teacher models with or without the further fine-tuning step produce similar student models.

\input{table1}

\textbf{TinyBERT and Simplified TinyBERT.}
For the general distillation, we use 3.5G raw text from English Wikipedia, where the 
losses for distilling the intermediate layer, namely,  $\mathcal{L}_{attn}$, $\mathcal{L}_{hidn}$, 
and $\mathcal{L}_{emb}$, are used.
The hyper-parameter temperature $T$ is fixed as 1 for both TinyBERT and the Simplified TinyBERT in the task-specific distillation stage, akin to the configuration in~\cite{DBLP:tinybert}.

\textbf{Training.}
The models are trained on up to four TITAN RTX 24G GPUs with Mixed Precision Training~\cite{DBLP:MicikeviciusNAD18}, using Adam optimizer with a weight decay of 0.01. In the general stage, we train for three epochs, setting learning rate 1e-6 and batch size 128.
In the task-specific stage, we perform distillation up to two epochs. 
We train with batch size equaling to 128, and learning rate to 1e-6 for Standard KD and the second step of TinyBERT, meanwhile using 64 and 5e-5 for Simplified TinyBERT and the first step of TinyBERT. 
We do model selection according to MRR@10 on validation set, apart from the general stage and the first step in the task-specific stage of TinyBERT, for which the last model is chosen for further distillation.


%% file: table1.tex
\begin{table*}
\centering
\caption{The results for different distilled models. L and H refer to the number of layers and the dimension of hidden states, respectively. Statistical significance at p-value $<$ 0.01 (0.05) is marked with $T(t)$ and $B(b)$ for comparisons to TinyBERT and the teacher model BERT-Base (L12\_H768), respectively. Note that MS MARCO Dev and TREC 2019 DL Test contain 4466 and 43 queries, respectively. }\label{tab:results}
\resizebox{\textwidth}{32mm}{
\begin{tabular}{l|cc|ccc|c}
\hline
\multirow{2}{*}{\textbf{Model (Size)}} & \multicolumn{2}{c}{\textbf{MS MARCO Dev}} & \multicolumn{3}{|c|}{\textbf{TREC 2019 DL Test}} & \textbf{FLOPs} \\ & \multicolumn{1}{c}{\textbf{MRR}} & \multicolumn{1}{c|}{\textbf{MRR@10}} & \multicolumn{1}{c}{\textbf{MRR}} & \multicolumn{1}{c}{\textbf{NDCG@10}} & \multicolumn{1}{c|}{\textbf{MAP}} & \textbf{(Speedup)} \\\hline

\textbf{L12\_H768} (109M) & 0.3589 & 0.3523  & 0.9341 & 0.6644 & 0.2861 & 22.9G (1$\times$) \\ \hline
\textbf{L6\_H768} (67M) & & & & & & \multirow{6}{*}{11.5G (2$\times$)}\\ 
Standard KD & $0.3570^T$ & $0.3498^T$ & 0.9341 & 0.6408 & 0.2783 \\
TinyBERT & $0.3711^B$ & $0.3646^B$ & 0.9380 & 0.6627 & 0.2821 \\
+ hard label  & $0.3767^{tB}$ & $0.3701^{tB}$ & 0.9380 & 0.6659 & 0.2777 \\
+ use one step & $0.3701^B$ & $0.3634^B$ & \textbf{0.9496} & 0.6620 & 0.2843 \\ 
Simplified TinyBERT & $\textbf{0.3908}^{TB}$ & $\textbf{0.3848}^{TB}$ & \textbf{0.9496}  & \textbf{0.6774} & \textbf{0.2847} \\ \hline

\textbf{L3\_H384} (17M) & & & & & & \multirow{6}{*}{1.5G (15$\times$)} \\ 
Standard KD & $0.3234^{TB}$ & $0.3148^{TB}$ & 0.9225 & $0.6042^B$ & $0.2567^B$ \\
TinyBERT & 0.3527 & 0.3453 & 0.8973 & $0.6230^b$ & 0.2755 \\
+ hard label & 0.3544 & 0.3470 & 0.9263 & $0.6361^{t}$ & $0.2721^B$ \\
+ use one step & $0.3630^T$ & $0.3560^T$ & 0.9263 & $0.6479^{t}$ & $0.2776^b$ \\ 
Simplified TinyBERT & $\textbf{0.3683}^{Tb}$ & $\textbf{0.3614}^{Tb}$ & \textbf{0.9554} & $\textbf{0.6698}^{T}$ & \textbf{0.2804} \\ \hline
\end{tabular}
}

\end{table*}

%% file: expResults.tex
\section{Experiments}\label{sec:exp}
\subsection{Experimental Setup}\label{sec:dataset}

\textbf{Dataset.}
According to our experiments, a relatively huge amount of training data is required to distill a small but effective BERT re-ranker. Meanwhile, recent work~\cite{DBLP:BertDistill} also demonstrates that about 5-10M training examples are required to distill a model that is comparable to BERT-Base on the passage ranking task. Thus, we employ MS MARCO document ranking dataset due to its largest available number of training samples, which contains 367,013 training queries, 5,193 development (dev) queries and 5,793 test queries (for leaderboard). In addition, TREC 2019 DL Track~\cite{DBLP:TREC2019DL} provides 43 test queries with more annotated relevant documents (compared with MS MARCO) based on the manual judgments from NIST assessors. We report our experiment results on the above two benchmarks.

\textbf{Data preprocessing.}
For training, after splitting the documents, we use the teacher model to filter passages from relevant documents, and reserve the five top-ranked passages as positive samples. 
Meanwhile, a negative passage is randomly sampled from irrelevant documents for every positive sample, to balance the positive and negative samples.
Thereby, the actual training set includes about 3.3M query-passage pairs.
For evaluation, due to the lack of annotations for test queries, we randomly reserve 727 dev queries for validation, and use the remaining 4,466 dev queries as our test set (but also denoted as MS MARCO Dev).
The max length of the input tokens in BERT re-ranker is set as 256. Statistical significance in terms of paired two-tailed t-test is reported.

\textbf{Models in comparison.}
The distilled models are compared under two configurations, namely, distilling BERT-Base ($\rm L12\_H768$) into a medium-size model ($\rm L6\_H768$) which provides 2$\times$ speedup relative to BERT-Base; and into a even smaller model ($\rm L3\_H384$) with only three layers, which provides 15$\times$ speedup. 
Several distillation models are included in Table~\ref{tab:results} for comparisons.
\textbf{Standard KD} distills the teacher model only using the prediction layer, namely, training the student model with $\alpha\mathcal{L}_{soft}+(1-\alpha)\mathcal{L}_{hard}$. We perform grid search on validation set over temperature $T$ = $\{1,5,10\}$ and $\alpha$ = $\{0.2,0.5,0.7\}$ on a parameters-fixed student model as in~\cite{DBLP:SunCGL19}; \textbf{TinyBERT} distills the teacher model following a two-stage method as in~\cite{DBLP:tinybert}, \textbf{Simplified TinyBERT} is the modified TinyBERT as described in Section~\ref{sec:modified_tinybert}.
In addition, \textbf{$+$ hard label} and \textbf{$+$ use one step} in Table~\ref{tab:results} indicate the results when applying one simplification on top of TinyBERT.

\subsection{Results}\label{sec:expResults}
In this section, we discuss the re-ranking performance of Standard KD, TinyBERT, and our Simplified TinyBERT.

\textbf{Distilled models perform well on document ranking task.} 
We first examine the performance of Standard KD and TinyBERT. 
For $\rm L6\_H768$, from Table~\ref{tab:results}, it can be seen that TinyBERT outperforms BERT-Base ($\rm L12\_H768$) significantly on our MS MARCO Dev set, and behaves on par with BERT-Base on TREC 2019 DL Test set, when providing 2$\times$ speedup.
For $\rm L3\_H384$, with 15$\times$ speedup, TinyBERT performs significantly worse than BERT-Base on shallow pool, and is comparable with BERT-Base on deep pool.
Compared with Standard KD, TinyBERT improves almost all metrics consistently, highlighting the strength of the distillation framework in TinyBERT.
Overall, according to our experiments, we confirm that both TinyBERT and Standard KD could dramatically reduce the model size meanwhile preserving most of the effectiveness.

\textbf{Simplified TinyBERT provides better re-ranking effectiveness and 15$\times$ speedup at the same time.}
We further examine the performance of the proposed Simplified TinyBERT, by comparing it with BERT-Base and TinyBERT.
From Table~\ref{tab:results}, on our MS MARCO Dev set, our Simplified TinyBERT could consistently outperform both BERT-Base and TinyBERT significantly under both model configurations.
On TREC 2019 DL Test set, when distilling a medium-size student model ($\rm L6\_H768$), Simplified TinyBERT performs on par with BERT-Base and TinyBERT; meanwhile, it outperforms TinyBERT on shallow pool in terms of NDCG@10, whereas TinyBERT performs significantly worse than BERT-Base, when the student model is very small ($\rm L3\_H384$).
\input{table2}

\input{table3}

\textbf{Robustness at different re-ranking depth.}
We also examine the effectiveness of the 3-layer student model ($\rm L3\_H384$) at different re-ranking depth, namely, top-10, 20, 50, and 100 documents.
As shown in Table~\ref{tab:doc_rank_depth}, the original TinyBERT behaves on par with BERT-Base, whereas our Simplified TinyBERT can outperform BERT-Base significantly at all re-ranking depth. This further confirms the superior effectiveness of the proposed simplifications.

\textbf{Ablation study on two simplifications.}
As shown in Table~\ref{tab:results}, 
both the simplifications could boost the metric scores, meanwhile two simplifications together gain even higher ranking performance. 
Thus, training using Equation (\ref{equ:soft_hard}) could bring significant boost, wherein both simplifications
contribute. 

\textbf{Simplified TinyBERT can be trained faster.}
As described in Section~\ref{sec:background}, the training of TinyBERT
is decomposed into two stages, and the second stage further includes two steps.
In our Simplified TinyBERT, as described in Section~\ref{sec:modified_tinybert},
we merge the two steps in the second stage. 
The training time of the second stage in the original TinyBERT and our Simplified TinyBERT is summarized in Table~\ref{tab:train_time}, where the proposed one-step simplification could save around 42-52\% training time. This is important when training on large datasets, like MS MARCO dataset used in this work.

%% file: table2.tex
\begin{table}
\renewcommand\arraystretch{1.05}
\centering
\caption{Re-ranking the documents at different depth using distilled L3\_H384 models. The MRR@10 on our MS MARCO Dev set (4466 queries) is reported. The superscripts for statistical significance test are the same as in Table \ref{tab:results}. } \label{tab:doc_rank_depth}
\setlength{\tabcolsep}{2mm}{\begin{tabular}{lccc}
\hline
Depth & L12\_H768 & TinyBERT & Simplified TinyBERT \\ \hline  
10 & 0.2896 & 0.2892 & $0.2970^{Tb}$ \\
20 & 0.3195 & 0.3188 & $0.3288^{TB}$ \\
50 & 0.3395 & 0.3359 & $0.3509^{TB}$ \\
100 & 0.3523 & 0.3453 & $0.3614^{Tb}$ \\ \hline
\end{tabular}}
\end{table}

%% file: table3.tex
\begin{table}
\renewcommand\arraystretch{1.05}
\centering
\caption{Training time of the second stage in TinyBERT and Simplified TinyBERT.}\label{tab:train_time} 
\setlength{\tabcolsep}{2mm}{\begin{tabular}{lcc}
\hline
Model & TinyBERT (two steps) & Simplified TinyBERT (one step) \\ \hline
L6\_H768 & 29.95h (2.08$\times$) & 14.37h (1$\times$) \\
L6\_H384 & 20.45h (1.81$\times$) & 11.30h (1$\times$)\\
L3\_H768 & 18.93h (1.88$\times$) & 10.05h (1$\times$)\\
L3\_H384 & 15.87h (1.72$\times$) & 9.22h (1$\times$)\\ \hline
\end{tabular}}
\end{table}

%% file: conclusions.tex
\section{Conclusion}\label{sec:conclusions}
In this paper, we demonstrated that the BERT-Base re-ranker model can be compressed using knowledge distillation technique, without compromising too much ranking effectiveness.
Furthermore, a simplified TinyBERT is proposed, the student model from whom could outperform the more expensive teacher model significantly. 
For the future work, we would like to study the distillation of more advanced ranking models like T5~\cite{DBLP:conf/emnlp/NogueiraJPL20} using the proposed knowledge distillation method.